\pdfoutput=1

\documentclass[conference]{IEEEtran}
\IEEEoverridecommandlockouts
\usepackage{cite}
\usepackage[T1]{fontenc}
\usepackage{amsmath,amssymb,amsfonts}
\usepackage{algorithmic}
\usepackage{graphicx}
\usepackage{textcomp}
\usepackage{xcolor}
\usepackage[hyphens,spaces,obeyspaces]{url}

\urldef{\footurl}\url{https://github.com/MTG/WGANSing}
\usepackage{caption}
\usepackage{float}
\usepackage{booktabs}
\usepackage{siunitx}
\begin{document}

\title{WGANSing: A Multi-Voice Singing Voice Synthesizer Based on the Wasserstein-GAN}

\author{\IEEEauthorblockN{Pritish Chandna\IEEEauthorrefmark{1},
Merlijn Blaauw\IEEEauthorrefmark{1}, Jordi Bonada\IEEEauthorrefmark{1} and
Emilia G\'omez\IEEEauthorrefmark{1}\IEEEauthorrefmark{2}}
\IEEEauthorblockA{\IEEEauthorrefmark{1}Music Technology Group, Universitat Pompeu Fabra, Barcelona, Spain\\
\IEEEauthorrefmark{2}Joint Research Centre, European Commission, Seville, Spain}\\
Email:pritish.chandna@upf.edu,
merlijn.blaauw@upf.edu,
jordi.bonada@upf.edu,
emilia.gomez@upf.edu}
\maketitle

\maketitle

\begin{abstract}
We present a deep neural network based singing voice synthesizer, inspired by the Deep Convolutions Generative Adversarial Networks (DCGAN) architecture and optimized using the Wasserstein-GAN algorithm. We use vocoder parameters for acoustic modelling, to separate the influence of pitch and timbre. This facilitates the modelling of the large variability of pitch in the singing voice. Our network takes a block of consecutive frame-wise linguistic and fundamental frequency features, along with global singer identity as input and outputs vocoder features, corresponding to the block of features. This block-wise approach, along with the training methodology allows us to model temporal dependencies within the features of the input block. For inference, sequential blocks are concatenated using an overlap-add procedure. We show that the performance of our model is competitive with regards to the state-of-the-art and the original sample using objective metrics and a subjective listening test. We also present examples of the synthesis on a supplementary website and the source code via GitHub.
\end{abstract}


\section{Introduction}
Singing voice synthesis and Text-To-Speech (TTS) synthesis are related but distinct research fields. While both fields try to generate signals mimicking the human voice, singing voice synthesis models a much higher range of pitches and vowel durations. In addition, while speech synthesis is controlled primarily by textual information such as words or syllables, singing voice synthesis is additionally guided by a score, which puts constraints on pitch and timing. These constraints and differences also cause singing voice synthesis models to deviate somewhat from their speech counterparts.
Historically, both speech and singing voice synthesis have been based on concatenative methods, which involve transformation and concatenation of waveforms from a large corpus of specialized recordings. Recently, several machine learning based methods have been proposed in both fields, most of which also require a large amount of data for training.  In terms of quality, the field of TTS has seen a revolution in the last few years, with the introduction of the WaveNet \cite{oord2016wavenet} autoregressive framework, capable of synthesizing speech virtually indistinguishable from a real voice recording. This architecture inspired the Neural Parametric Singing Synthesiser (NPSS) \cite{blaauw2017neural}, a deep learning based singing voice synthesis method which is trained on a dataset of annotated natural singing and produces high quality synthesis.

The WaveNet \cite{oord2016wavenet} directly generates the waveform given local linguistic and global speaker identity conditions. While a high quality synthesis is generated, the drawback of this model is that it requires a large amount of annotated data. As such, some following works, like the Tacotron 2 \cite{shen2018natural}, use the WaveNet as a vocoder for converting acoustic features to a waveform and use a separate architecture for modelling these acoustic features from the linguistic input. The WaveNet vocoder architecture, trained on unlabeled data, is also capable of synthesizing high-quality speech from an intermediate feature representation. The task that we focus on in this paper is generating acoustic features given an input of linguistic features. 

Various acoustic feature representations have been proposed for speech synthesis, including the mel-spectrogram \cite{shen2018natural}, which is a compressed version of the linear-spectrogram. However, for the singing voice, a good option is to use vocoder features, as they separate pitch from timbre of the signal. This is ideal for the singing voice as the pitch range of the voice while singing is much higher than that while speaking normally. Modelling the timbre independently of the pitch has been shown to be an effective methodology \cite{blaauw2017neural}. We note that the use of a vocoder for direct synthesis can lead to a degradation of sound quality, but this degradation can be mitigated by the use of a WaveNet vocoder trained to synthesize the waveform from the parametric vocoder features. As such, for the scope of this study, we limit the upper-bound of the performance of the model to that of the vocoder. 

Like autoregressive networks, Generative adversarial networks (GANs) \cite{goodfellow2014generative,mirza2014conditional,salimans2016improved} are a family of generative frameworks for deep learning, which includes the Wasserstein-GAN \cite{arjovsky2017wasserstein} variant. While the methodology has provided exceptional results in fields related to computer vision, it has only a few adaptations in the audio domain and indeed in TTS, that we discuss in the following sections. We adapt the Wasserstein-GAN model for singing voice synthesis. In this paper, we present a novel block-wise generative model for singing voice synthesis, trained using the Wasserstein-GAN framework\footnote{The code for this model in the TensorFlow framework is available at \mbox{\footurl}. Audio examples, including synthesis with voice change and without the reconstruction loss can be heard at \mbox{\url{https://pc2752.github.io/sing_synth_examples/}}}. The block-wise nature of the model allows us to model temporal dependencies among features, much like the inter-pixel dependencies are modelled by the GAN architecture in image generation. For this study, we use the original fundamental frequency for synthesis, leading to a performance driven synthesis. As a result, we only model the timbre of the singing voice and not the expression via the $f0$ curve or the timing deviations of the singers. 


The rest of the paper is organized as follows. Section \ref{sec:GANs} provides a brief overview of the GAN and Wasserstein-GAN generative frameworks. Section \ref{sec:related} discusses the state-of-the-art singing voice synthesis model that we use as a baseline in this paper and some of the recent applications of GANs in the field of TTS and in general, in the audio domain. The following sections, section \ref{sec:system} and section \ref{sec:feats} present our model for singing voice synthesis, followed by a brief discussion on the dataset used and the hyperparameters of the model in sections \ref{sec:data} and \ref{sec:hyper} respectively. We then present an evaluation of the  model, compared to the baseline in section \ref{sec:results}, before wrapping up with the conclusions of the paper and a discussion of our future direction in section \ref{sec:conclusion}.

\section{GANs And Wasserstein-GANs}
\label{sec:GANs}
Generative Adversarial Networks (GANs) have been extensively used for various applications in computer vision since their introduction. GANs can be viewed as a network optimization methodology based on a two-player non-cooperative training that tries to minimize the divergence between a parameterized generated distribution $P_g$ and a real data distribution, $P_r$. It consists of two networks, a generator, $G$ and a discriminator, $D$, which are trained simultaneously. The discriminator is trained to distinguish between a real input and a synthetic input output by the generator, while the generator is trained to fool the discriminator. The loss function for the network is shown in equation \ref{eq:GAN}.

\begin{equation} 
\begin{aligned}[b]
\mathcal{L}_{GAN} =  \min\limits_{G}\max\limits_{D} \mathbb{E}_{y\sim P_r} [\log(D(y))]\\ + \mathbb{E}_{x\sim P_x} [\log(1-D(G(x)))] 
\end{aligned}
\label{eq:GAN}
\end{equation} 
Where $y$ is a sample from the real distribution and $x$ is the input to the generator, which may be noise or conditioning as in the Conditional GAN \cite{mirza2014conditional} and is taken from a distribution of such inputs, $P_x$.

While GANs have been shown to produce realistic images, there are difficulties in training including vanishing gradient, mode collapse and instability. To mitigate these difficulties, the Wasserstein-GAN \cite{arjovsky2017wasserstein} has been proposed, which optimizes an efficient approximation of the Earth-Mover (EM) distance between the generated and real distributions and has been shown to produce realistic images. The loss function for the WGAN is shown in equation \ref{eq:WGAN}. In this version of the GAN, the discriminator network is replaced by a network termed as critic, also represented by $D$, which can be trained to optimality and does not saturate, converging to a linear function. 

\begin{equation} 
\mathcal{L}_{WGAN} = \min\limits_{G}\max\limits_{D} \mathbb{E}_{y\sim P_r} [D(y)] - \mathbb{E}_{x\sim P_x} [D(G(x))] 
\label{eq:WGAN}
\end{equation} 

We use a conditional version of the model, which generates a distribution, parametrized by the network and conditioned on a conditional vector, described in section \ref{sec:feats} and follow the training algorithm proposed in the original paper \cite{arjovsky2017wasserstein}, with the same hyperparameters for training. 

\section{Related Work}
\label{sec:related}

GANs have been adapted for TTS in several variations over recent years. The work closest to ours was the one proposed by Zhao et al. \cite{zhao2018wasserstein}, which uses a Wasserstein-GAN framework, followed by a WaveNet vocoder and a complimentary waveform based loss. Yang et al.  \cite{yang2017statistical} use the mean squared error (MSE) and a variational autoencoder (VAE) to enable the GAN optimization process in a multi-task learning framework. A BLSTM based GAN framework complemented with a style and reconstruction loss is used by Zhao et al. \cite{ma2018a}. While these models use recurrent networks for frame-wise sequential prediction, we propose a convolutional network based system to directly predict a block of vocoder features, based on an input conditioning of the same size in the time dimension. 

Other examples of the application of GANs for speech synthesis include work done by Kaneko et al., \cite{kaneko2017generative} and  \cite{kaneko2017generative1}, which use GANs as a post-filter for acoustic models to overcome the oversmoothing related to the models used. 
GANs have also been adapted to synthesize waveforms directly; WaveGAN \cite{donahue2018adversarial} is an example of the use of GANs to synthesize spoken instances of numerical digits, as well as other audio examples. GANSynth \cite{engel2018gansynth} has also been proposed to synthesize high quality musical audio using GANs.

For singing voice synthesis, Hono et al. \cite{hono2019singing} use a GAN-based architecture to model frame-wise vocoder features. This models the inter-feature dependencies within a frame of the output. In contrast, our model directly models a block of consecutive audio frames via the Wasserstein-GAN framework. This allows us to model temporal dependencies between features within that block. This temporal dependence is modelled via autoregression in the Neural Parametric Singing Synthesizer (NPSS) \cite{blaauw2017neural} model, which we use as a baseline in our study.

The NPSS uses an autoregressive architecture, inspired by the WaveNet \cite{oord2016wavenet}, to make frame-wise predictions of vocoder features, using a mixture density output. The network models the frame-wise distribution as a sum of Gaussians, the parameters of which are estimated by the network. In contrast, the use of adversarial networks for estimation, imposes no explicit constraints on the output distribution. The NPSS model has been shown to generate high quality singing voice synthesis, comparable or exceeding state-of-the-art concatenative methods. A multi-singer variation of the NPSS model has also been proposed recently \cite{blaauw2019data}, and is used as the baseline for our study.

\begin{figure}[H]
\centering
\includegraphics[width=\linewidth]{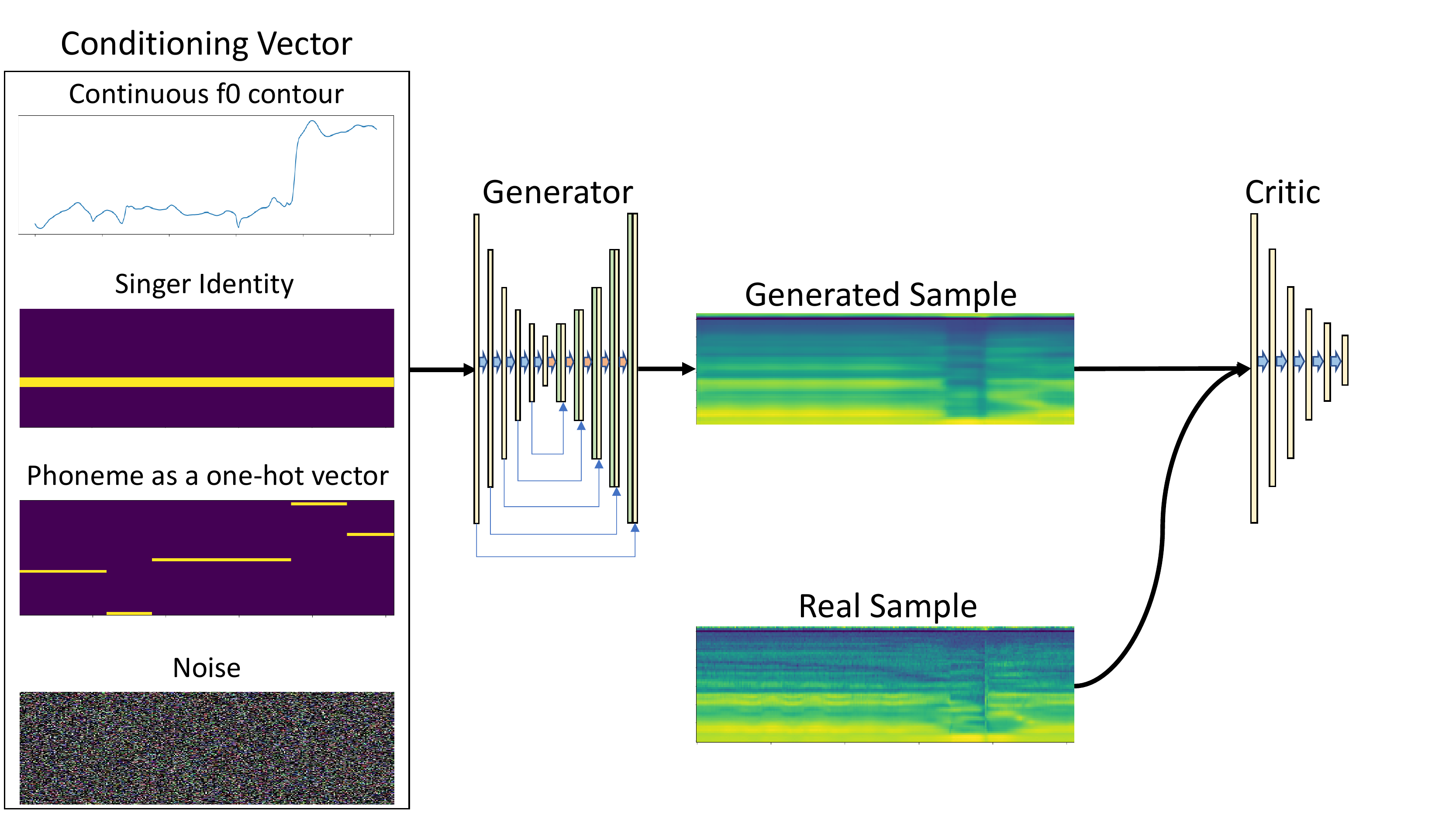}
 \caption{The framework for the proposed model. A conditioning vector, consisting of frame-wise phoneme and $f0$ annotations along with speaker identity is passed to the generator. The critic is trained to distinguish between the generated sample and a real sample.}
 \label{fig:framework}
\end{figure}  
\section{Proposed System}
\label{sec:system}

\begin{figure}[H]
\centering
\includegraphics[width=\linewidth]{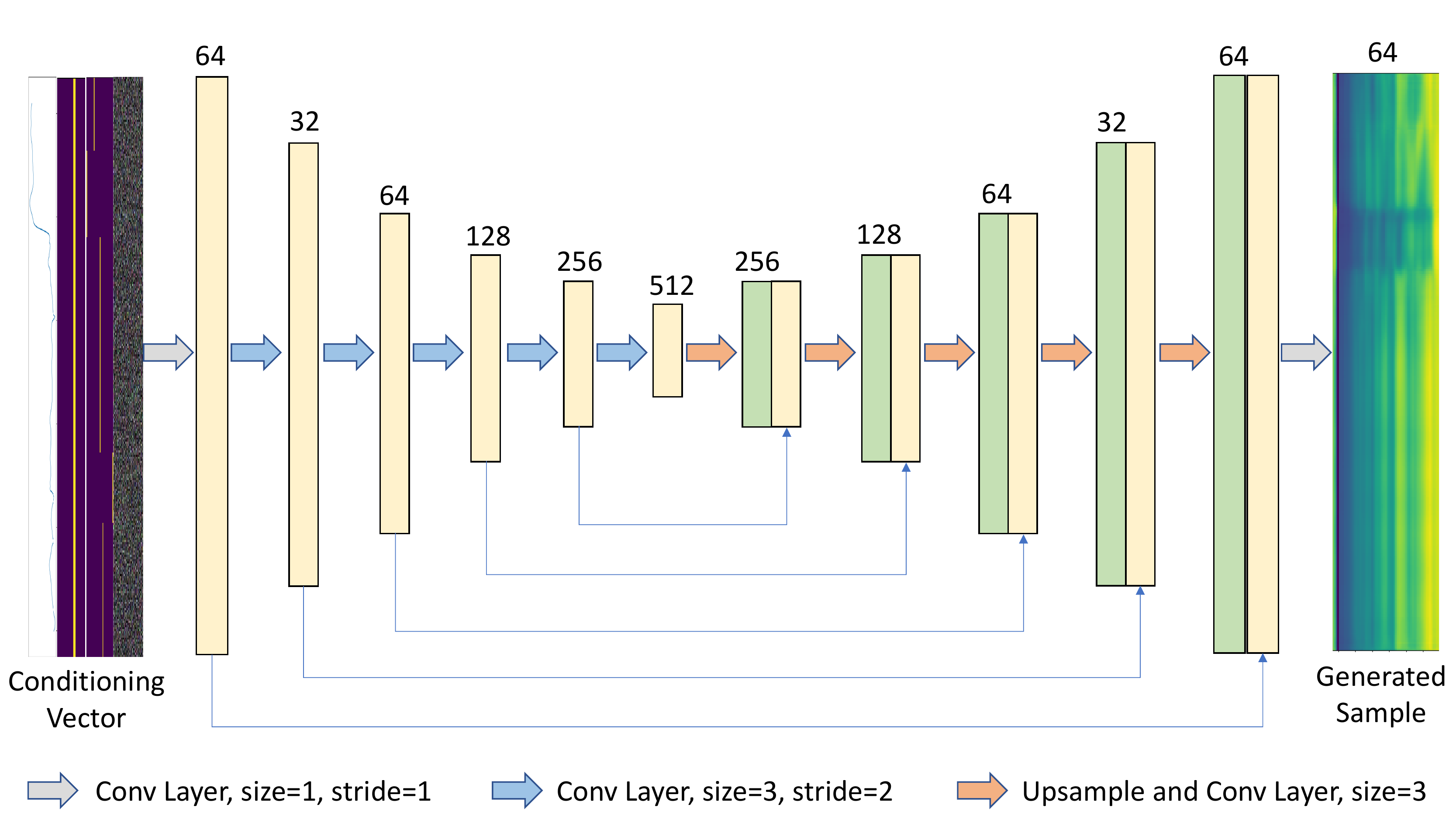}
 \caption{The architecture for the generator of the proposed network. The generator consists of an encoder and a decoder, based on the U-Net architecture \cite{ronneberger2015u}.}
 \label{fig:generator}
\end{figure}

We adopt an architecture similar to the DCGAN \cite{radford2015unsupervised}, which was used for the original WGAN. For the generator, we use an encoder-decoder schema, shown in figure \ref{fig:generator} wherein both the encoder and decoder consist of $5$ convolutional layers with filter size $3$. Connections between the corresponding layers of the encoder and decoder, as in the U-Net \cite{ronneberger2015u} architecture are also implemented. 

\begin{figure}[H]
\includegraphics[width=0.45\textwidth]{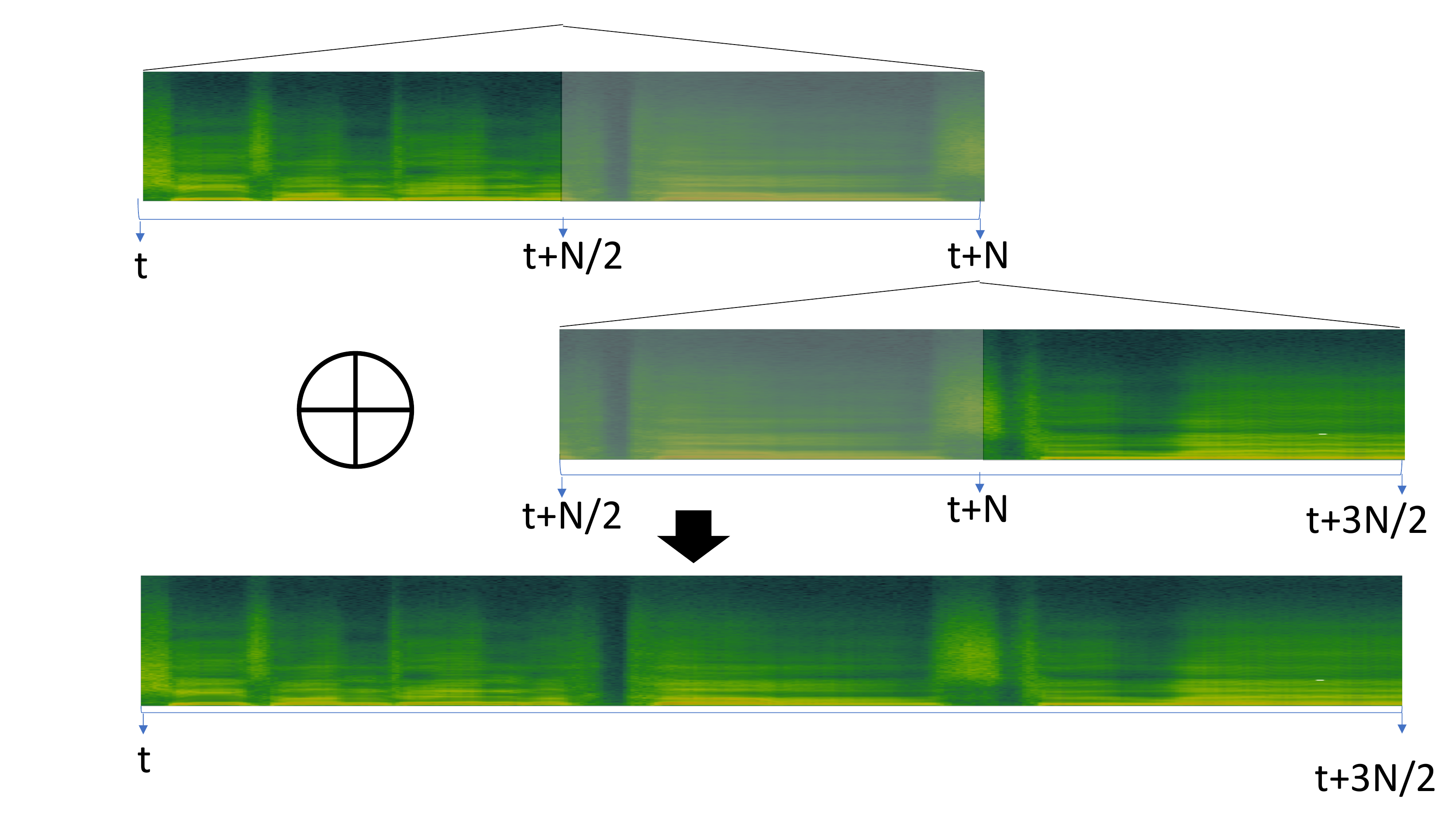}
 \caption{The overlap add process for the generated features. As shown, predicted features from time $t$ to time $t+N$ are overlap-added with features from time $t+N/2$ to $t+3N/2$, where $t$ is the start time of the process in view. A triangular window is used for the adding process, applied across each of the features.}
 \label{fig:overlap}
\end{figure}

While convolutional networks are capable of modelling arbitrary length sequences, the critic in our model takes a block of a fixed length input, thereby modelling the dependencies within the block. Furthermore, the encoder-decoder schema leads to conditional dependence between the features of the generator output, within the predicted block of features. This approach implies implicit dependence between vocoder features of a single block but not within the blocks themselves. As a result, for inference, we use overlap-add of consecutive blocks of output vocoder features, as shown in figure \ref{fig:overlap}. An overlap of $50\%$ was used with a triangular window across features. This appraoch is similar to that used by Chandna et al. \cite{chandna2017monoaural}
As proposed by Radford et al. \cite{radford2015unsupervised}, we use strided convolutions in the encoder instead of deterministic pooling functions for downsampling. For the decoder, we use linear interpolation followed by normal convolution for upsampling instead of transposed convolutions, as this has been shown to avoid the high frequency artifacts which can be introduced by the latter \cite{stoller2018wave}. Blocks of size $N$ consecutive frames are passed as input to the network and the output has the same size. 
Like the DCGAN, we use ReLU activations for all layers in the generator, except the final layer, which uses a tanh activation. We found that the use of batch normalization did not affect the performance much. We also found it helpful to guide the WGAN training by adding a reconstruction loss, as shown in equations \ref{eq:L1} and \ref{eq:WL1}. This reconstruction loss is often used in conditional image generation models like Lee et al.\cite{lee2018harmonizing}. 
\begin{gather}
\mathcal{L}_{recon} = \min\limits_{G}\mathbb{E}_{x,y} \left\lVert G(x)-y\right\rVert
\label{eq:L1}\\
 \mathcal{L}_{total} = \mathcal{L}_{WGAN} + \lambda_{recon} \mathcal{L}_{recon}
\label{eq:WL1}
\end{gather}
Where $\lambda_{recon}$ is the weight given to the reconstruction loss. The networks are optimized following the scheme described in Arjovsky et al.\cite{arjovsky2017wasserstein}.
The critic for our system uses an architecture similar to the encoder part of the generator, but uses LeakyReLU activation instead of ReLU, as used by Radford et al. \cite{radford2015unsupervised}. 

Convolutional neural networks offer translation invariance across the dimensions convolved, making them highly useful in image modelling. However, for audio signals, this invariance is useful only across the time-dimension but undesirable across the frequency dimension. As such, we follow the approach of NPSS \cite{blaauw2017neural}, representing the features as a 1D signal with multiple channels.


 

\section{Linguistic And Vocoder Features}
\label{sec:feats}
The input conditioning to our system consists of frame-wise phoneme annotations, represented as a one-hot vector and continuous fundamental frequency extracted by the spectral autocorrelation (SAC) algorithm. This conditioning is similar to the one used in NPSS. In addition, we condition the system on the singer identity, as a one-hot vector, broadcast throughout the time dimension. This approach is similar to that used in the WaveNet \cite{oord2016wavenet}. The three conditioning vectors are then passed through a $1\times1$ convolution and concatenated together along with noise sampled from a uniform distribution and passed to the generator as input. 

We use the WORLD vocoder \cite{morise2016world} for acoustic modelling of the singing voice. The system decomposes a speech signal into the harmonic spectral envelope and aperiodicity envelope, based on the fundamental frequency $f0$. We apply dimensionality reduction to the vocoder features, similar to that used in \cite{blaauw2017neural}. 


\section{Dataset}
\label{sec:data}
We use the NUS-48E corpus \cite{duan2013nus}, which consists of $48$ popular English songs, sung by $12$ male and female singers. The singers are all non-professional and non-native English speakers. Each singer sings $4$ different songs from a set of $20$ songs, leading to a total of \SI{169}{\minute} of recordings, with \num{25474} phoneme annotations. We train the system using all but $2$ of the song instances, which are used for evaluation.

\section{Hyperparameters}
\label{sec:hyper}
A hoptime of \SI{5}{\milli\second} was used for extracting the vocoder features and the conditioning. We tried different block-sizes, but empirically found that $N=128$ frames, corresponding to \SI{640}{\milli\second} produced the best results.

We used a weight of $\lambda_{recon} = 0.0005$ for $\mathcal{L}_{recon}$ and trained the network for \num{3000} epochs. As suggested in \cite{arjovsky2017wasserstein}, we used RMSProp for network optimization, with a learning rate of $0.0001$.
After dimension reduction, we used \num{60} harmonic and \num{4} aperiodic features per frame, leading to a total of \num{64} vocoder features. 

\section{Evaluation Methodology}
For objective evaluation, we use the Mel-Cepstral Distortion metric. This metric is presented in table \ref{table:mcd}.
For subjective evaluation, we used an online AB test wherein the participants were asked to choose between two presented \SIrange{5}{7}{\second} examples\footnote{We found that WGANSing without the reconstruction loss as a guide did not produce very pleasant results and did not include this in the evaluation. However, examples for the same can be heard at \mbox{\url{https://pc2752.github.io/sing_synth_examples/}}}, representing phrases from the songs. The participant's choice was based on the criteria of Intelligibility and Audio Quality. We compare our system to the NPSS, trained on the same dataset. Along with the NPSS, we use a re-synthesis with the WORLD vocoder as the baseline as this is the upper limit of the performance of our system. We compared $3$ pairs for this evaluation:
\begin{itemize}
    \item WGANSing - Original song re-synthesized with WORLD vocoder.
    \item WGANSing - NPSS
    \item WGANSing, original singer - WGANSing, sample with different singer.

\end{itemize}

\begin{figure}[H]
 
\includegraphics[width=\linewidth]{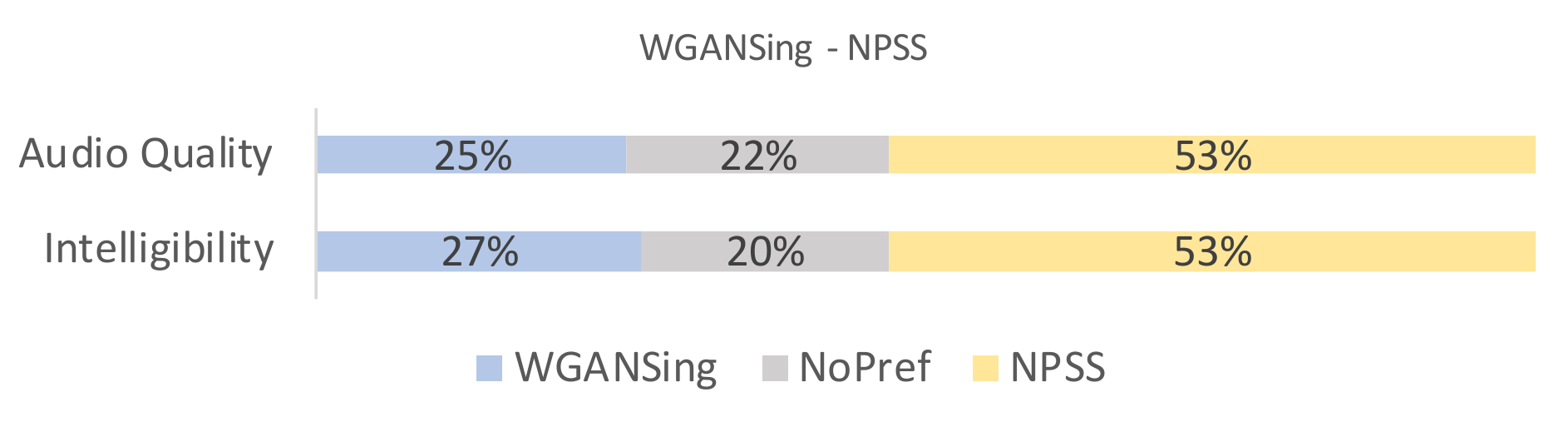}
 \caption{Subjective test results for the WGANSing-NPSS pair.}
 \label{fig:ganpss}
\end{figure} 

For the synthesis with a changed singer, we included samples with both singers of the same gender as the original singer and of a different gender. The input $f0$ to the system was adjusted by an octave to account for the different ranges of the genders. For each criteria, the participants were presented with $5$ questions for each of the pairs, leading to a total of $15$ questions per criteria and $30$ questions overall\footnote{The subjective listening test used in our study can be found at \mbox{\url{https://trompa-mtg.upf.edu/synth_eval/}}}.

\section{Results}
\label{sec:results}

\begin{figure}[H]
 
\includegraphics[width=\linewidth]{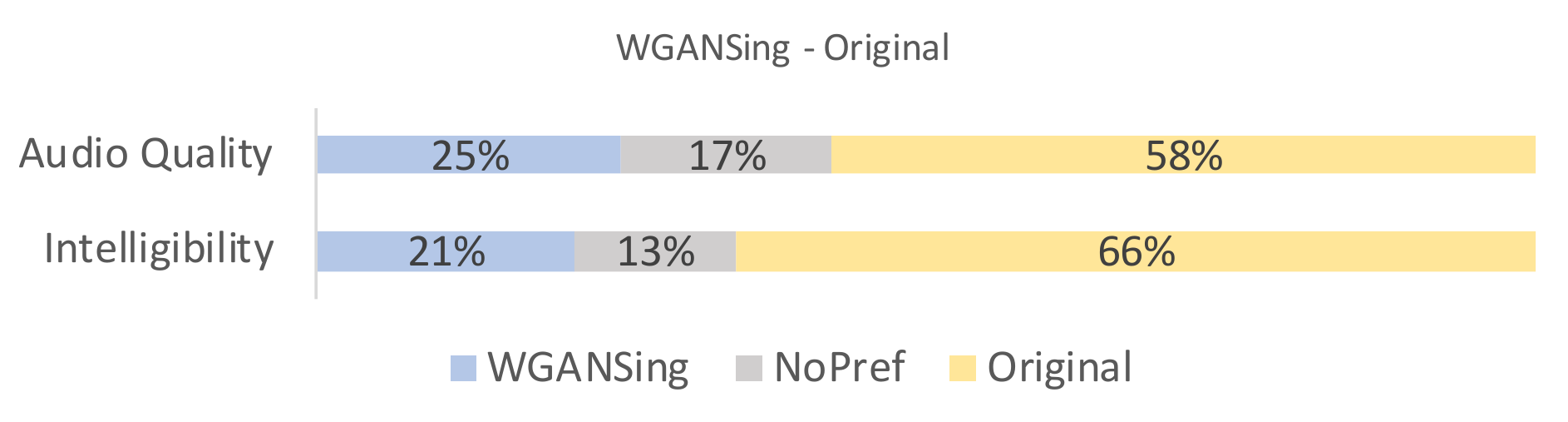}
 \caption{Subjective test results for the WGANSing-Original pair.}
 \label{fig:ganori}
\end{figure}
\begin{figure}[H]
\includegraphics[width=\linewidth]{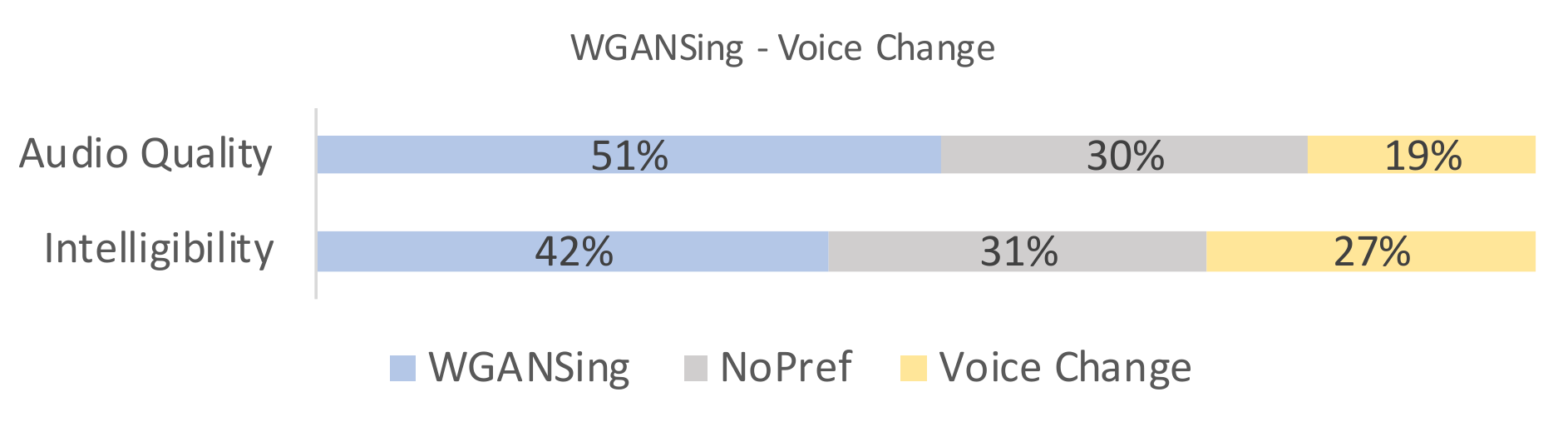}
 \caption{Subjective test results for the WGANSing-WGANSing Voice Change pair.}
 \label{fig:gancha}
\end{figure}

There were a total of $27$ participants from over $10$ nationalities, including native English speaking countries like the USA and England, and ages ranging from $18$ to $37$ in our study. The results of the tests are shown in figures \ref{fig:ganori}, \ref{fig:ganpss} and \ref{fig:gancha}. 

From the first two figures, it can be seen that our model is qualitatively competitive with regards to both the original baseline and the NPSS, even though a preference is observed for the later. This result is supported by the objective measures, seen in table \ref{table:mcd}, which show parity between WGANSing and the NPSS models. Figure \ref{fig:gancha} shows that the perceived intelligibility of the audio is preserved even after speaker change, even though there is a slight compromise on the audio quality.
Variability in the observed results can be attributed to the subjective nature of the listening test, the diversity of participants and the dataset used, which comprises of non-native, non-professional singers. We note that there is room for improvement in the quality of the system, as discussed in next section.
\begin{table}[H]\centering
\begin{tabular}{@{}llll@{}}
\toprule
Song           & WGAN + $\mathcal{L}_{recon}$   & WGAN        & NPSS        \\
\midrule
Song 1 JLEE 05 & 5.36 dB & 9.70 dB & 5.62 dB \\
Song 2 MCUR 04 & 5.40 dB & 9.63 dB & 5.79 dB \\
\bottomrule
\end{tabular}

\caption{The MCD metric for the two songs used for validation of the model. The three models compared are the NPSS\cite{blaauw2017neural} and the WGANsing model with and without the reconstruction loss.}
\label{table:mcd}
\end{table}

 

\section{Conclusions And Discussion}
\label{sec:conclusion}
We have presented a multi-singer singing voice synthesizer based on a block-wise prediction topology. This block-wise methodology, inspired by the work done in image generation allows the framework to model inter-block feature dependency, while long-term dependencies are handled via an overlap-add procedure. We believe that synthesis quality can further be improved by adding the previously predicted block of features as a further conditioning to current batch of features to be predicted. Further contextual conditioning, such as phoneme duration, which is currently implicitly modelled, can also be added to improve synthesis. Synthesis quality can also be improved through post-processing techniques such as the use of the WaveNet vocoder on the generated features. 


We also note that the synthesis was greatly helped by the addition of the reconstruction loss, while the use of batch normalization did not affect performance either way. The synthesis quality of the model was evaluated to be comparable to that of stat-of-the-art synthesis systems, however, we note that variability in subjective measures like intelligibility and quality is introduced during the listening test, owing to the diversity of the population participating. This variability can be reduced through a targeted listening test with expert participants. The generative methodology used allows for potential exploration in expressive singing synthesis. Furthermore, the fully convolutional nature of the model leads to faster inference than autoregressive or recurrent network based models. 



\section*{acknowledgments}
This work is partially supported by the European Commission under the TROMPA project (H2020 770376). The TITAN X used for this research was donated by the NVIDIA Corporation.
\vspace{12pt}
\bibliographystyle{IEEEtran}
\bibliography{biblo}

\end{document}